\documentclass[twocolumn,showpacs]{revtex4}
\usepackage{amssymb}
\usepackage{graphicx}
\usepackage{dcolumn}

\def\be{\begin{equation}}
\def\ee{\end{equation}}
\def\bea{\begin{eqnarray}}
\def\eea{\end{eqnarray}}
\input epsf
\begin{document}

\title{{\Large
On the new string theory inspired mechanism of generation\\
of  cosmological perturbations
}}

\author{Nemanja Kaloper$^{1}$ Lev Kofman,$^2$ Andrei Linde$^{3,4}$ and Viatcheslav Mukhanov$^4$}
\affiliation{$^1$Department of Physics, University
of California, Davis, CA 95616, USA\\
$^2$ CITA, University of Toronto, 60 St George Str, Toronto, ON
M5S 3H8, Canada\\
$^{3}$Physics Department, Stanford University, Stanford CA 94305-4060, USA\\
$^{4}$ASC, Physics \ Department, LMU, Theresienstr. 37, Munich,
Germany}

\begin{abstract}
Recently a non-inflationary mechanism of generation of scale-free cosmological perturbations of metric was proposed by Brandenberger, Nayeri,  and Vafa in the context of the string gas cosmology.  We discuss various problems of their model and argue that the cosmological perturbations of metric produced in this model have blue spectrum with a spectral index $n = 5$, which strongly disagrees with observations. We conclude that this model in its present form is not a viable alternative to inflationary cosmology.
\end{abstract}

\pacs{98.80.Cq } \maketitle

\section{Introduction}

At present, inflation is the only well established mechanism which
solves the homogeneity, isotropy, flatness and horizon problems,
and explains why the universe has such a large mass and entropy.
In addition, it provides a simple mechanism of generation of
metric perturbations of with a flat spectrum, as required for the formation of large-scale structure of the
universe. For a review of inflationary theory and the theory of
perturbations of metric there see e.g.  \cite{book,mukhbook}.

There were many attempts to suggest an alternative solution to the major cosmological problems, but so far the progress in this direction was quite limited. For example,
the pre-big bang scenario \cite{PBB} and the ekpyrotic/cyclic
scenario \cite{cyclic} are based on the assumption that eventually
one will find a solution of the cosmological singularity problem
and learn how one could transfer small perturbations of metric
through the singularity. This problem, as well as several other problems discussed in
\cite{nogo,KKL,Linde:2002ws}, still remain unsolved. An
alternative
explanation of the origin of cosmological perturbations was
advocated in \cite{Hollands:2002yb}.
The mechanism was based on the alternation of the QFT rules.
However this mechanism does not
address any other cosmological problems, and it relies on
speculations about
dynamics of perturbations  at an epoch when the energy density was
$10^{95}$ times greater than the Planck density
\cite{Kofman:2002cj}. In comparison, inflation solves the major
cosmological problems and produces
metric perturbations with a
flat spectrum practically independently of the speculative
processes at extremely high energies and large curvatures, or before the big
bang.

In a recent series of papers
\cite{Nayeri:2005ck,Brandenberger:2006xi,Nayeri,Brandenberger:2006vv}
it was claimed that after a certain modification of the string theory inspired
cosmological model
proposed in \cite{Brandenberger:1988aj,Tseytlin:1991xk} one can
obtain scale free cosmological perturbations of metric due to the
physical processes {\it after} the big bang. This could be  advantageous over the pre-big bang and the ekpyrotic/cyclic
scenario, because in those models the cosmological perturbations are
produced {\it before} the big bang, making the corresponding conclusions
unreliable. The authors of \cite{Nayeri:2005ck,Brandenberger:2006xi,Nayeri,Brandenberger:2006vv}
admitted that their model does not solve the flatness problem and
has several other unexplained features. Thus it is not a real alternative to inflation. Nevertheless it would be quite interesting to have a   non-inflationary mechanism of production of perturbations of metric with a flat spectrum.
Our goal here is to examine this issue.

In Section 2 we will go over some of the properties and unsolved problems of the basic cosmological model of \cite{Nayeri:2005ck,Brandenberger:2006xi,Nayeri,Brandenberger:2006vv}.
In Section 3 we will argue that the spectrum of scalar perturbations produced in this model is dramatically tilted. Instead of the scale-free spectrum with the spectral index $n = 1$, the perturbations produced in the model of \cite{Nayeri:2005ck,Brandenberger:2006xi,Nayeri,Brandenberger:2006vv} have blue spectrum with the spectral index $n = 5$, disagreeing with the cosmological data. The main problem
with the proposal of \cite{Nayeri:2005ck,Brandenberger:2006xi,Nayeri,Brandenberger:2006vv} stems from the fact that they worked in the string frame, but directly employed the standard
description of metric perturbations \cite{mukhbook} tailored for
the Einstein frame. Meanwhile, due to a nontrivial dilaton dynamics these two frames strongly differ from each other at the epoch when the perturbations of metric were produced. In Section 4 we will show that a similar conclusion is  valid for the tensor perturbations of metric as well (though for a different reason): the
spectrum of the gravitational waves produced in their scenario is far from being flat.

\section{Basic scenario }\label{basic}

\subsection{Tseytlin-Vafa solution}\label{TV}

The cosmological scenario of Ref.
\cite{Nayeri:2005ck,Brandenberger:2006xi} is based in part on the
cosmological solution obtained by Vafa and Tseytlin
\cite{Tseytlin:1991xk}. This solution describes an $N$-dimensional
universe in the context of the effective dilaton gravity theory
with the action
\be\label{action} S =  {1\over 2} \int d^{N+1}x~ \sqrt{-G}~
e^{{-2\phi}} \left(R - 4(D\phi)^{2}\right) \, . \ee
Here $G$ is the metric determinant of the $N+1$ dimensional space,
$\phi$ is the dilaton field, $e^{{\phi}} = g_{s} = ({ M_{s}/
M_{P}})^{{(N-1)/2}}$ is string coupling. In addition to the
gravitational degrees of freedom of action Eq. (1),
\cite{Nayeri:2005ck,Brandenberger:2006xi} also include the
contribution of a gas of strings following its thermodynamic
description explained in \cite{Tseytlin:1991xk}. An important role
is
purportedly played by winding modes of strings and p-branes, as
discussed, e.g., in \cite{Brandenberger:2001kj,Battefeld:2005av}

The string frame metric
\be ds^{2}_{\rm S} = dt^{2} -a^{2}(t) d^{2}\vec x \, , \ee
is related to the Einstein frame metric by a conformal transformation
\be \label{eframe}
ds^{2}_{\rm E} = \exp \left(-{4\phi\over
N-1}\right)\, ds^{2}_{\rm S} = d\tau^{2} -\alpha^{2}(\tau)
d^{2}\vec x \ . \ee
One should emphasize that once one replaces string theory by the effective action for metric and the dilaton field
of the type of (\ref{action}), as was done in
\cite{Tseytlin:1991xk,Nayeri:2005ck,Brandenberger:2006xi,Nayeri},
all calculations can be equally well performed either in the
string frame, or in the Einstein frame, where the standard Einstein equations are valid. The final results describing physical quantities should be
equivalent, independently of the frame used to compute them in, as
long as the correct conformal transformation rules are used on the
metric and throughout the matter sector, including any possible
$UV$ regulators. In other words, the domain of validity of the Einstein frame and Einstein equations coincides with the domain of validity of the effective string theory action (\ref{action}).

Usually string theory calculations are performed in the string
frame
\cite{Tseytlin:1991xk,Nayeri:2005ck,Brandenberger:2006xi,Nayeri}.
An important
convenience of the Einstein frame is that the gravitational part
of the effective action (\ref{action}) in this frame looks like
the standard Einstein action with the time-independent
gravitational constant. The methodology   and physical picture
for the investigation of metric perturbations has been
originally developed in the Einstein frame, where one can use the
standard, simple Einstein equations. When the dilaton is constant,
the two frames coincide with each other, up to an overall
normalization coefficient. However, in general, the dilaton is not
constant, and can sometimes lead to an apparition of a profoundly different picture of cosmological
evolution as seen from these two frames. Yet, however, as we have already stressed above, the differences
are only apparent, since the physical quantities remain
independent of the conformal frame.

Consider, for example, the only explicit cosmological solution
obtained in \cite{Tseytlin:1991xk} (the last equation in the
Appendix). It describes an isotropic expansion in all N
directions:
\be\label{VT1} e^{-\varphi} ={Et^2\over 4}-{NA^2\over E} \ , \ \
\lambda =\lambda_0+{1\over {\sqrt N}}{\rm
ln}{t-2{\sqrt{N}}A/E\over t+2\sqrt{N} A/E} \, . \ee
Here $A$ and $E$ are some constants, the relation between $\phi$
and $\varphi$ is given by
\be e^{\phi} = g_{s} = \Bigl({ M_{s}\over
M_{P}}\Bigr)^{\frac{N-1}{2}} =e^{\varphi/2} \, e^{N\lambda/2} =
e^{\varphi/2} a^{N/2} \, , \ee
and the scale factor $a(t)$ of the spatially flat $FRW$ metric
with $N$ spatial dimensions is given by $a(t) = e^{\lambda(t)}$.

After a change of variables, $t \to t -t_{0}/2$, with
$t_{0}={4\sqrt N A\over E}$, this solution can be represented in a
more convenient form:
\be e^{-\varphi} =  {E\over 4}|t| (|t|+t_{0})\ , \quad \lambda =
\lambda_{0}+{ 1\over \sqrt N} \log {|t|+t_{0}\over |t|}\, . \ee
It describes the evolution of the universe when $t$ grows from
$-\infty$ to $0$.

The scale factor is given by
\be\label{scale} a = e^{\lambda_{0}} \left({|t| + t_{0} \over |t|}\right)^{1/\sqrt N}  \ . \ee
When $t$ grows from $-\infty$ to $0$, the size of the universe
{\it increases}, and it {\it blows up} near $t = 0$ as
$|t|^{-1/\sqrt N}$, in string  units of length, $l_{s} =
M_{s}^{-1}$. Note that in the beginning of the evolution of the
universe, at $t \to -\infty$, the universe was static, $a =
e^{\lambda_{0}}$. One
might at first wonder how something like this could ever happen,
because the usual Einstein equations for a flat universe imply
\be\label{einst} H_{E}^{2} =  {\rho\over 3 M_{P}^{2}} \, , \ee
where $H_E = \frac{1}{\alpha} \frac{d\alpha}{d\tau}$ is the
Einstein frame Hubble constant. Thus the universe can be static only if density of matter
vanishes, which is not the case considered in
\cite{Tseytlin:1991xk}.

 However
 equations in  \cite{Tseytlin:1991xk} have been solved in
 string frame, where they look quite different from
the Einstein equation (\ref{einst}) due to the nontrivial dilaton behavior. Interestingly, in the
Einstein frame the same cosmological evolution looks not as a
cosmological expansion (\ref{scale}), but rather as a {\it
gravitational collapse}. One can easily show that whereas the size
of the universe $a(t)$ in the string frame {\it grows}, the same size
measured in units of $M_{P}^{{-1}}$, i.e. the scale factor in the
Einstein frame $\alpha(\tau)$,  {\it decreases}:
\bea \alpha  \sim |t|^{2/(N-1)}\sim  {|\tau|}^{2/(N+1)}  \, , \eea
where $\tau$ the Einstein frame time, as defined in Eq.
(\ref{eframe}).

 This example vividly illustrates why one should be extremely
careful with specifying the particular frame used in the
investigation of string cosmology. While the results of the
calculations in each frame are  of course physically equivalent,
one may run into all kinds of paradoxical conclusions by directly
applying equations valid in one frame to another frame. In
particular, one should not use the Einstein equations for the
description of dynamics in the string frame. Meanwhile, as we will see, this is
exactly what was done in the computation of cosmological
perturbations in
\cite{Nayeri:2005ck,Brandenberger:2006xi,Nayeri,Brandenberger:2006vv},
where no distinction between these two frames was made.

 We should stress a fact that will be crucial for us in what
follows. The transition between $a(t)$ and $\alpha(\tau)$ is
achieved by the simple rescaling corresponding to the change of
units of length from string units to the Planck units:
\be\label{scalefacts} \alpha = e^{-{2\phi\over N-1}} \, a = {M_P\over M_{s}}\, a \,
. \ee
Meanwhile the relation between the Hubble
constant in the string frame $H_{s} = {\dot a \over a}$, where the overdot denotes string frame
time derivative, and the usual Hubble constant in the Einstein frame $H_{E} = {{d\alpha\over
d\tau}\over\alpha}$ is more involved:
\be H_{E} = e^{2\phi\over N-1}\left(H_{s}- {2 \dot \phi \over
N-1}\right) = {M_{s}\over M_{P}}\left(H_{s}- {2\dot\phi\over
N-1}\right) \, . \ee
We will return to this point in Section \ref{NBVpert}.

\subsection{ The model of Brandenberger, Nayeri  and Vafa}\label{NBV}

The model of
\cite{Nayeri:2005ck,Brandenberger:2006xi,Nayeri,Brandenberger:2006vv}
is only partially related to the Tseytlin-Vafa model. The
description of the basic cosmological scenario in
\cite{Nayeri:2005ck,Brandenberger:2006xi,Nayeri,Brandenberger:2006vv}
is rather incomplete. It contains many  speculative
elements, which makes it very hard to analyze. We review the basic issues here.

1) Tseytlin-Vafa solution discussed in the previous section
described isotropic expansion in all $N$ spatial dimensions.
Meanwhile Refs.
\cite{Nayeri:2005ck,Brandenberger:2006xi,Nayeri,Brandenberger:2006vv}
consider a different possibility and assume that only three
dimensions are expanding, while the others are stabilized. This assumption was intensely
debated in the literature on string gas cosmology. A detailed
investigation of
\cite{Easther:2004sd,Danos:2004jz} found no support for this hypothesis. Recently a novel mechanism to address this problem was proposed, involving
$RR$ flux in unwrapped dimensions \cite{Kim:2006yx}, but this
mechanism relies on the dilaton stabilization, which is another as
yet unsolved problem of this scenario, see below.

2) The three expanding dimensions are supposed to be extremely
large from the very beginning, of a size greater than $10^{23}$ in
string units. It is not quite clear how this could be related to
the original ideas of the T-dual description of Ref.
\cite{Brandenberger:1988aj}, where the initial size of all
dimensions was supposed to be $O(1)$ in string units. The authors
acknowledge this as a manifestation of the flatness problem of their new scenario. This problem is more general, however, because the general
asymmetric and anisotropic initial data generally will not evolve
towards the homogeneous and isotropic initial state needed for
\cite{Nayeri:2005ck,Brandenberger:2006xi,Nayeri,Brandenberger:2006vv},
see \cite{Tseytlin:1991xk}.

3) In addition to the flatness problem, one should also solve the
homogeneity problem. A mechanism addressing this problem was proposed in
\cite{Brandenberger:2005qj}. It was based on the introduction of a
specific potential between two orbifold fixed planes. Such potential does not follow from string theory, and the authors
acknowledged that this proposal is extremely speculative
\cite{BrandInfl25}. On the other hand, a slight modification of
one of the assumptions of Ref. \cite{Brandenberger:2005qj} leads
to inflation \cite{Shuhmaher:2005pw}, which makes alternative
mechanisms of generation of perturbations of metric unnecessary.

4) Just like the Tseytlin-Vafa solution \cite{Tseytlin:1991xk}
described above, the scenario of
\cite{Nayeri:2005ck,Brandenberger:2006xi,Nayeri,Brandenberger:2006vv}
is based on the assumption that initially the universe was static,
so that the size of the horizon $H_{s}^{-1} = {a\over \dot a}$ was
infinite. However, by the word ``initially'' the authors of
\cite{Nayeri:2005ck,Brandenberger:2006xi,Nayeri,Brandenberger:2006vv}
do not mean $t \to -\infty$. Instead, they either simply assume, without providing specific details,  that at some intermediate stage of expansion the universe was static but slowly accelerating, or mention a specific loitering regime  \cite{BrandInfl25} when the universe, according to
\cite{Brandenberger:2001kj}, should have experienced several stages of expansion and
contraction, in the string frame. The ``initial''
static universe in the loitering regime should correspond to the last bounce, after which the universe should switch to a more regular stage of expansion.

Dynamics of the loitering regime begs for an additional investigation. In particular, according to
\cite{Brandenberger:2001kj}, the  transition from the loitering
phase should lead to a singular behavior of all fields, unless the
dilaton field is stabilized by the dilaton potential $V(\phi)$,
which should survive even at the present stage of the evolution of
the universe.
Most importantly,
stabilization of the dilaton, as well as of many other string
theory moduli fields, is
necessitated by particle phenomenology and by the theory of
gravitational interactions, specifically by the equivalence
principle tests, in order to yield acceptable low energy limits
which conform with observations.

The problem of
stabilizing the dilaton and other string theory moduli was studied extensively for the last 20 years, but it proved to
be extremely complicated. A possible solution to this problem was
proposed only very recently, in the context of the KKLT
construction \cite{Giddings:2001yu,Kachru:2003aw}. However, this
construction involves many additional ingredients, such as branes,
fluxes etc., which are not a part of the model considered in
\cite{Nayeri:2005ck,Brandenberger:2006xi,Nayeri,Brandenberger:2006vv}.

5) The scenario of \cite{Nayeri:2005ck,Brandenberger:2006xi,Nayeri,Brandenberger:2006vv}   assumes the existence of a thermal equilibrium in the Hagedorn phase. An investigation of this issue in \cite{Danos:2004jz} suggests that no such equilibrium can be maintained. A similar conclusion was reached in  \cite{Takamizu:2006sy}, where it was argued that the dilaton stabilization by a hypothetical dilaton potential is required for solving this problem. But if the dilaton is stabilized during the Hagedorn phase,  this  completely invalidates the mechanism of generation of density perturbations proposed in \cite{Nayeri:2005ck,Brandenberger:2006xi,Nayeri,Brandenberger:2006vv}: This mechanism is based on the rapid shrinking of the Hubble radius $H_{s}^{-1}$. In the models with the stabilized dilaton, this assumption contradicts the null energy condition, see Sections \ref{para}.

The absence of many essential ingredients makes it rather
difficult to discuss the evolution of the universe and generation of cosmological perturbations in the scenario of
\cite{Nayeri:2005ck,Brandenberger:2006xi,Nayeri,Brandenberger:2006vv}.
Nevertheless we will try to follow the logic of Refs.
\cite{Nayeri:2005ck,Brandenberger:2006xi,Nayeri,Brandenberger:2006vv},
keeping all of our reservations in mind. We will attempt to evaluate whether this scenario, if it can work despite all of the problems discussed above, may provide a
new mechanism of generation of scale invariant metric
perturbations.

\section{Density Perturbations}\label{NBVpert}

\subsection{Generation of scalar perturbations of metric}\label{NBVbasic}

Perhaps the simplest way to grasp the main idea of the mechanism
of generation of metric perturbations \cite{Nayeri:2005ck,Nayeri}
is to use the figure given in \cite{Nayeri:2005ck}, see Fig. 1 of
our paper. We added to their figure only one line, DB (dashed green
line). According to \cite{BrandInfl25}, this line, combined with
the line BC, describes the evolution of the Hubble horizon
$H_{E}^{-1}$ in the Einstein frame.

As we have already mentioned, the the relevant cosmological evolution described by this scenario
is taken to begin not at $t = -\infty$, but at some intermediate
stage, when the universe was static and the size of the Hubble
radius in the string frame $H_{s}^{{-1}}$ was infinite. Then the
universe begins to expand, and $H_{s}^{{-1}}$ starts to shrink.
The last
stage of the shrinking,
shown by the line AB,
describes the evolution of the quasi-static Hagedorn regime into
the radiation-dominated phase of standard cosmology. According to
\cite{Nayeri:2005ck}, this transition is supposed to occur very quickly; within a finite time the size
of the Hubble radius should shrink from an infinite value to $M_P/M_{s}^{2}$.
The main idea of Ref. \cite{Nayeri:2005ck,Nayeri} is that when the
Hubble radius $H_{s}^{-1}$ becomes smaller than the wavelength of
perturbations of metric created by thermal fluctuations inside the
horizon, these perturbations freeze, just like the metric perturbations do during inflation.

\begin{figure}[h!]
\centerline{\epsfxsize=3.7in\epsfbox{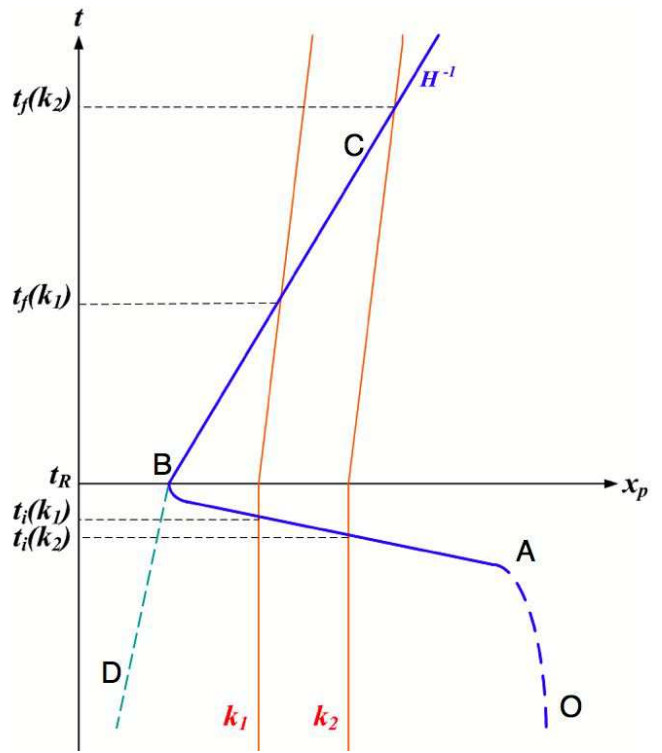}}
\caption{\small This figure illustrates the scenario outlined in
\cite{Nayeri:2005ck}. The vertical axis is time, the horizontal
axis is physical distance. The Hagedorn phase ends
at the time $t_R$ and is followed by the radiation-dominated phase
of standard cosmology. The blue curve represents the Hubble radius
$H_{s}^{-1}$ which is cosmological during the quasi-static
Hagedorn phase, shrinks abruptly to a microphysical scale at $t_R$
and then increases linearly in time for $t > t_R$. All scales here are given in
string  frame.  The only exception is the segment DB (dashed green
line). According to   \cite{BrandInfl25}, it describes the
evolution of the size of the horizon $H_{E}^{-1}$ in the Einstein
frame (rescaled by the factor of $M_{s}/M_{P}$) before the time of the radiation domination. Fixed comoving
scales (labeled by $k_1$ and $k_2$) which are currently probed in
cosmological observations have wavelengths which are {\it smaller} than
the string frame Hubble radius $H_{s}^{-1}$ during the Hagedorn phase. They exit the
Hubble radius $H_{s}^{-1}$ shown by the line AB at times $t_i(k)$ just prior to $t_R$, and propagate
with a wavelength larger than $H_{s}^{-1}$ until they reenter
the Hubble radius at times $t_f(k)$. Note, however,  that the wavelength of
 these perturbations  was many orders of magnitude {\it greater} than the
 Einstein frame horizon $H_{E}^{-1}$ at the time when they were generated,
 and they {\it did not} cross the Einstein frame horizon DB.} \label{fig:1}
\end{figure}

To describe this effect, one can use the longitudinal gauge,
where the perturbed metric  takes the form
\be\label{long} ds^2 \, = \, - (1 + 2 \Phi) dt^2 +  a^{2}(t) (1 - 2 \Phi) d{\bf
x}^2 \, . \ee
Here  $\Phi({\bf x}, t)$ represents the fluctuation mode.
The spectral index $n$ of the cosmological perturbations is
determined by
\be\label{spectrum} P_{\Phi}(k) \, \equiv \, k^3 |\Phi(k)|^2 \,
\sim k^{n - 1} \, . \ee
Here $\Phi(k)$ is the Fourier coefficient of $\Phi$ and $P_X$
denotes the dimensionless power spectrum of some quantity $X$. The
value $n = 1$ corresponds to a scale-invariant spectrum.

In order to calculate the amplitude of perturbations of metric,
Nayeri {\it et al} calculate the amplitude of thermodynamical
fluctuations of the energy density $\delta \rho$ in the Hagedorn
regime, and relate these perturbations to the metric perturbation $\Phi$  via the Einstein constraint equation
\be \label{constr} \nabla^2 \Phi \, = \, 4 \pi G \delta \rho \, .
\ee
Recall again that the standard Einstein equations are valid only
in the Einstein frame. Furthermore, this particular equation is
valid only for the sub-horizon perturbations with wavelengths
$\lambda < H_{E}^{{-1}}$. This fact will be important for us
later, but for the time being we will ignore this issue and
continue along the lines of
\cite{Nayeri:2005ck,Brandenberger:2006xi,Nayeri,Brandenberger:2006vv}.
The calculations of
\cite{Nayeri:2005ck,Brandenberger:2006xi,Nayeri,Brandenberger:2006vv}
imply that the spectrum of density perturbations is proportional
to $k^{4}$:
\be  \label{blue} P_{\rho} = k^{3}\left({\delta\rho(k)\over
\rho}\right)^{2} \sim k^{4} = k^{n_{\rho}-1} \, , \ee
with $n_{\rho} = 5$. According to
\cite{Nayeri:2005ck,Nayeri,Brandenberger:2006vv}, this spectrum
becomes transformed to the spectrum of  perturbations of metric
$\Phi$  with flat spectrum, $n = 1$, because the Einstein equation
(\ref{constr}) implies that $k^{2 }\Phi \sim \delta\rho$. Their
final result for the spectrum of metric fluctuations
\be \label{spectrum2} P_{\Phi} \, \sim \, 16 \pi^2 G^2
\alpha'^{-3/2} \frac{T}{1 - T/T_H} \, , \ee
which would imply that the spectrum of metric perturbations {\it
on scales smaller than the Hubble radius} is approximately
scale-invariant, $n = 1$. Then one can argue
\cite{Nayeri:2005ck,Brandenberger:2006xi,Nayeri,Brandenberger:2006vv}
that these perturbations freeze out in the interval AB, Fig. 1,
when the Hubble radius $H_{s}^{-1}$ becomes smaller than the
wavelength of the perturbations $k^{{-1}} $\footnote{If these
considerations were valid, then one could also infer that in the
loitering scenario of Ref. \cite{Brandenberger:2001kj} the
universe may have experienced several stages when the speed of
expansion of the universe in the string frame vanishes. Each time,
the Hubble radius $H_{s}^{-1}$ changes from some finite value to
an infinitely large one, and then becomes finite again. Following
the logic of Ref. \cite{Nayeri:2005ck,Nayeri,Brandenberger:2006vv}
one would then have expected that scale-invariant metric
perturbations could have been produced not only once, but several
times during the history of the universe in the string gas
cosmology.}. According to
\cite{Nayeri:2005ck,Nayeri,Brandenberger:2006vv}, this results in
generation of adiabatic perturbations of metric with flat
spectrum, without any need of inflation.

\subsection{Does it work?}\label{para}

At the first glance, the ideas described in the previous section
may seem quite plausible. However, as we have repeatedly noted,
investigation performed in
\cite{Nayeri:2005ck,Brandenberger:2006xi,Nayeri,Brandenberger:2006vv}
made no distinction between the string frame and the Einstein
frame. On the one hand, the evolution of the universe and the
behavior of all thermodynamic quantities have been studied in
\cite{Nayeri:2005ck,Brandenberger:2006xi,Nayeri,Brandenberger:2006vv}
in the string frame. On the other hand, the evolution of density
perturbations was studied using the Einstein equation
(\ref{constr}). This equation is valid only in the Einstein frame,
for the perturbations with the wavelength smaller than the
cosmological horizon evaluated in the Einstein frame. The string
frame and the Einstein frame do not differ much after the end of
the Hagedorn phase, at $t > t_{R}$, because the dilaton field at
this stage (line BC) moves very slowly
\cite{Tseytlin:1991xk,dilrad}. However the difference between
these two frames is very significant during the Hagedorn phase $t<
t_{R}$, and especially at the transition stage AB (stage DB in the
Einstein frame),  when the large-scale perturbations of metric
were generated.

With this in mind, let us first examine the assumption that the
perturbations of metric freeze when their wavelength becomes
greater than the Hubble radius.
The immediate question which then arises is, which Hubble radius
should we compare the perturbation wavelength to? The size of the
Einstein frame cosmological horizon $H_{E}^{{-1}}$ is {\it not}
equal to the rescaled value of $H_{s}^{-1}$:
\be H_{E}^{-1} =  {M_{P}\over M_{s}}\left(H_{s}- {2\dot\phi\over
N-1}\right)^{-1}   \ee
Up to the rescaling factor ${M_{P}\over M_{s}} = e^{-{2\phi\over
N-1}}$, the quantities $H_{E}^{-1}$ and $H_{s}^{-1}$  would
coincide for $\dot \phi = 0$. However, in the model of
\cite{Nayeri:2005ck,Brandenberger:2006xi,Nayeri,Brandenberger:2006vv}
one has $\dot\phi <0$. Therefore the (rescaled) Einstein frame
Hubble radius is {\it smaller} than the Hubble radius in the
string frame. In particular, the universe in the Einstein frame
continues to expand and the horizon $H_{E}^{-1} $ remains finite
even it appears to stop in the sting frame, when $H_{s} = 0$ and
the string frame Hubble radius $H_{s}^{-1}$ is infinite.

Of course, if $\dot\phi$ is small (e.g. after the onset of
radiation domination, line BC \cite{Tseytlin:1991xk,dilrad}),
then, up to the rescaling, there is no much difference between
$H_{s}^{{-1}}$ and $H_{E}^{{-1}}$. This is shown in Fig. 1 by the
line BC. It simultaneously describes $H_{s}^{{-1}}$ and the
rescaled horizon $H_{E}^{{-1}}$, because the rescaling factor
${M_{P}\over M_{s}}$ remains constant along this line as the
dilaton does not move at that time. On the other hand, as we noted
above, these two quantities differ a lot during the most important
stage AB, where, according to
\cite{Nayeri:2005ck,Brandenberger:2006xi,Nayeri,Brandenberger:2006vv},
the perturbations of metric with the wavelengths greater than
Hubble radius are supposed to be frozen.

As noted in \cite{Tseytlin:1991xk}, this transition from the
Hagedorn phase to radiation cosmology is only partially
understood. Investigation of related subjects in \cite{Jain} was
made without taking into account the evolution of the dilaton and
the background metric, which as we saw are quite significant
during this stage. Indeed, if the dilaton were frozen during this
stage, then there would be no difference between the string frame
and the Einstein frame, and the size of the horizon could only
grow, according to the conventional Einstein equations, which
yield
\begin{equation}\label{fried2}
\dot H_{E}  = - 4\pi G (\rho + p)  \, .
\end{equation}
This quantity cannot be positive unless one studies phantom matter
with $\rho + p<0$, which violates the null energy condition and
typically leads to a catastrophic vacuum instability
\cite{Carroll:2003st}. Thus whenever the null energy condition is
satisfied, the size of the horizon in the Einstein frame can only
grow. This is shown in Fig. 1, in accordance with
\cite{BrandInfl25}. At the stage AB the string frame
$H_{s}^{{-1}}$ shrinks, while at the same time $H_{E}^{{-1}}$
expands, as shown by line DB. The difference between these two
regimes is possible only because of the rapid dilaton evolution at
that time.

It is important to realize that the difference between the
rescaled values of $H_{E}^{{-1}}$ and $H_{s}^{{-1}}$ at this stage
is extremely large. Indeed, according to Fig. 1, the rescaled
value of $H_{E}^{{-1}}$ along the line DB is smaller than the
minimal value $H_{s}^{{-1}} \sim M_{P}/M_{s}^{2}$ on the line AB.
On the other hand, if we want the perturbations generated during
that epoch to describe the present large scale structure of the
universe, the value of $H_{s}^{{-1}}$ at the time when the seeds
for large-scale perturbations are generated must be exponentially
larger than the minimal value of $H_{s}^{{-1}} \sim
M_{P}/M_{s}^{2}$. Thus we are talking about the situation where
the difference between the rescaled value of $H_{E}^{{-1}}$ and
$H_{s}^{{-1}}$ is {\it exponentially} large.

Given such a profound difference between $H_{E}^{{-1}}$ and
$H_{s}^{{-1}}$ at the stage when the metric perturbations are
generated, one should check exactly when they freeze, and in fact
if they freeze at all. Does it happen when their wavelength
becomes greater than $H_{E}^{{-1}}$, or when it becomes greater
than  $H_{s}^{{-1}}$? Freezing of the oscillations is a
frame-invariant effect, so one should be able to analyze it in any
frame. However, the exploration of this issue in the string frame
is rather complicated because perturbations of the metric interact
not only with gravity but also with the dilaton. Moreover,
$H_{s}^{{-1}}$ does not have any direct relation to particle
horizon, event horizon, or apparent horizon. For example, when
$H_{s}$ vanishes and $H_{s}^{{-1}}$ becomes infinite, the universe
does not instantly become causally connected. Thus, there is no
obvious reason to assume that fluctuations should freeze when
their wavelength becomes greater than $H_{s}^{{-1}}$.

Fortunately, to resolve this issue one can study freezing out of
the the metric perturbations in the Einstein frame, and the answer
is well known: The perturbations freeze if their wavelength is
greater than the Hubble radius in the Einstein frame
$H_{E}^{{-1}}$, rather than at the moment when their wavelength
becomes greater than  $H_{s}^{{-1}}$, as assumed in
\cite{Nayeri:2005ck,Nayeri,Brandenberger:2006vv}.

But as we just mentioned, the Hubble radius in the Einstein frame
can never shrink, unlike the Hubble radius in the string frame.
This fact is illustrated by Fig. 1, which shows that while
$H_{s}^{-1}$ shrinks in the interval AB, the size of the horizon
in the Einstein frame $H_{E}^{{-1}}$ grows in the corresponding
interval DB. Since $H_{E}^{{-1}}$ can only grow, unless the null
energy condition is violated, one may wonder whether it is at all
possible for the perturbations with a given comoving momentum
initially greater than the Einstein frame Hubble scale $H_E$ to go
outside the horizon and freeze there?

The answer to this question is also well known. If the universe
experiences power law expansion $\alpha(\tau) \sim \tau^{p}$ with
$p < 1$, which is the typical situation in non-inflationary
cosmology, then the horizon always expands faster that the
wavelengths, $H_{E} \sim \tau$, and the sub-horizon fluctuations
never freeze. On the other hand, if the size of the horizon for a
long time practically does not change, then the wavelengths
eventually become greater than $H_{E}^{{-1}}$ and the fluctuations
can freeze. This is the standard inflationary mechanism. But the
authors
\cite{Nayeri:2005ck,Brandenberger:2006xi,Nayeri,Brandenberger:2006vv}
assume that there is no inflation in their scenario. Then we are
forced to conclude that no freeze-out of the fluctuations occurs
when $H_{s}^{-1}$ suddenly shrinks to a very small value along the
line AB.

This does not necessarily preclude production of metric
perturbations at this time. Indeed, if thermal fluctuations of
density exist at that stage, they certainly induce some metric
perturbations. However, there are two problems here. First of all,
whereas the long-wavelength perturbations discussed in
\cite{Nayeri:2005ck,Nayeri,Brandenberger:2006vv} were crossing the
shrinking radius $H_{s}^{-1}$, their wavelength was always
exponentially greater than the horizon $H_{E}^{{-1}}$, as we just
discussed. It is not obvious to us what mechanism could be
responsible for the establishment of the thermal equilibrium
between the fluctuations on scales which are exponentially greater
than the size of the horizon; see also general comments about the
applicability of thermal description in this regime in
\cite{Takamizu:2006sy}. On the other hand, if the long-wavelength
modes are out of thermal equilibrium, then the calculations of
density perturbations performed in
\cite{Nayeri:2005ck,Nayeri,Brandenberger:2006vv} simply cannot
remain valid at the cosmologically large, super-horizon, scales.

But let us ignore this issue for a moment and assume, following
\cite{Nayeri:2005ck,Nayeri,Brandenberger:2006vv}, that  the
spectral distribution of density perturbations is $P_{\rho} \sim
k^{4} = k^{n_{\rho}-1}$ with $n_{\rho} = 5$, see Eq. (\ref{blue}).
As we already mentioned, \cite{Nayeri:2005ck,
Nayeri,Brandenberger:2006vv} assert that this spectrum becomes
transformed into the spectrum of metric perturbations $\Phi$ with
flat spectrum, $n = 1$, because the Einstein equation
(\ref{constr}) implies that $k^{2 }\Phi \sim \delta\rho$. But as
it stands, this equation is valid only for perturbations {\it with
the wavelengths smaller than the horizon $H_{E}^{{-1}}$} (not
$H_{s}^{{-1}}$!).

For the perturbations with the wavelengths greater than the
horizon $H_{E}^{{-1}}$, the relation between $\Phi$ and
$\delta\rho$ in the longitudinal gauge (\ref{long}) used in
\cite{Nayeri:2005ck,Nayeri,Brandenberger:2006vv} is quite
different  \cite{mukhbook}:
\be\label{outside} 2\Phi(k) = -\, {\delta \rho(k)\over \rho} \, .
\ee
This equation remains true for {\it any} relation between the
wavelengths of the perturbations and $H_{s}^{{-1}}$, as long as
these wavelengths are greater than $H_{E}^{{-1}}$. Thus it is valid
 for all long-wavelength cosmological perturbations we are
interested in, see Fig. 1. This implies, according to
(\ref{blue}),  that perturbations outside the horizon have the
same spectrum as ${\delta \rho(k)\over \rho}$, i.e. instead of the
flat spectrum with $n = 1$ they have a blue spectrum
\be  \label{blue2} P_{\Phi}   \sim k^{4} = k^{n-1} \, , \ee
which, with the approximations inherent in
\cite{Nayeri:2005ck,Nayeri,Brandenberger:2006vv}, yields $n =
n_{\rho}= 5$, in contrast to their claim that $n = 1$. This
spectrum is ruled out by cosmological observations.

Furthermore, suppose, for example, that the amplitude of
perturbations of metric and density on the horizon scale
$H_{E}^{-1}$   is equal to $10^{{-5}}$. One can then easily check
using Eq. (\ref{blue2}) that the relative perturbations of mass
${\delta M \over M}$ on a given scale will become much greater
than $O(1)$ for $k > 10^{3} H_{E}$. This means that the
perturbation theory used in \cite{Nayeri:2005ck,
Nayeri,Brandenberger:2006vv} can cover no more than 3 orders of
magnitude of scale, which is inadequate for describing
cosmological perturbations in our universe.

Finally, let us note that if thermodynamical equilibrium is
somehow maintained during the transition from the Hagedorn phase
to the radiation-dominated phase of standard cosmology, then the
amplitude of perturbations of metric always follows the evolution
of the perturbations of density, in accordance with Eq.
(\ref{outside}). But during this transition, the perturbations of
density in the Hagedorn phase gradually evolve into the usual
thermodynamical perturbations of density in the
radiation-dominated universe, which can never produce large scale
cosmological perturbations with flat spectrum.

The main difference between this situation and the situation in
inflationary cosmology can be explained as follows. In
inflationary cosmology the inflaton field is {\it not} in a state
of thermal equilibrium. Its fluctuations are frozen at the time
when their wavelength becomes greater than $H_{E}^{-1}$. These
frozen fluctuations serve as a source for the metric fluctuations,
which are also frozen in accordance with Eq. (\ref{outside}).

If we want to have something similar in the string gas scenario,
we would need to assume that the large scale perturbations of
stringy matter density are frozen. If this is so, they cannot be
in thermal equilibrium, because the properties of the universe in
thermal equilibrium change dramatically during the transitional
period. On the other hand, if the perturbations are not in thermal
equilibrium, then the calculations of their amplitude
\cite{Nayeri:2005ck, Nayeri,Brandenberger:2006vv} are not valid.
If we nevertheless assume that these calculations are valid, then,
as we have seen, the resulting metric perturbations have a blue
spectrum, with $n = 5$. Either way, we do not obtain scale free
cosmological perturbations in this scenario.

Does this exhaust all possibilities? First of all, until now, we
have been operating under the assumption made in
\cite{Nayeri:2005ck, Nayeri,Brandenberger:2006vv} that one can
calculate the amplitude of density perturbations in the string
frame and then simply plug in the result into the Einstein
equations. However, the amplitude of density perturbations in the
string frame is different from the amplitude of the perturbations
in the Einstein frame, where the Einstein equations can be used,
because of the dilaton contribution.

In the string frame, one could safely ignore this contribution
because in the Hagedorn phase the dilaton was just one out of many
degrees of freedom contributing to the energy density of the
universe. On the other hand, upon the conformal transformation to
the Einstein frame, which involves $e^{-{2\phi\over N-1}}$,  the
dilaton perturbations induce perturbations of the energy density
of all other fields. As a result one finds the following
expression relating the amplitude of density perturbations in the
Einstein frame and in the string frame \cite{Kalara:1990ar}:
\be\label{dilpert} \left({\delta\rho\over\rho}\right)_{E} =
\left({\delta\rho\over\rho}\right)_{S} +{8 \over N-1} \,
\delta\phi\ . \ee
Therefore to obtain the amplitude of metric perturbations from the
amplitude of perturbations of energy density calculated in the
string frame, one should write Eq. (\ref{outside}) more
accurately:
\be\label{outside2} 2\Phi(k) = -
\left({\delta\rho\over\rho}\right)_{E} = -
\left({\delta\rho\over\rho}\right)_{S} -{8 \over N-1} \,
\delta\phi\ . \ee
The term $\left({\delta\rho\over\rho}\right)_{S}$ was calculated
in \cite{Nayeri:2005ck, Nayeri,Brandenberger:2006vv} and, as we
argued above, it produces metric perturbations with blue spectrum,
which rapidly disappear in the large scale limit. If the dilaton
perturbations are thermal, then the dilaton contributions to
metric perturbations also have a blue spectrum with the amplitude
which rapidly disappears at the cosmological scale.

There is only one exception from this rule that we are aware of:
If the universe experiences a stage of inflation, and the dilaton
field remains very light during inflation, $H_{E} \gg m_{\phi}$,
then the standard inflationary dilaton perturbations with a flat
spectrum are produced. This results in the scale
invariant metric perturbations
\be\label{outside3} \Phi(k) =  -{4 \over N-1} \, \delta\phi\ .
\ee
This mechanism  is equivalent to the mechanism of generation of
modulated perturbations of metric  developed in \cite{mod1}.
However, this mechanism requires inflation.

Finally, if one is willing to consider even very exotic
possibilities to save the scenario of
\cite{Nayeri:2005ck,Brandenberger:2006xi,Nayeri,Brandenberger:2006vv},
one may look more attentively at our main premise that the
behavior of  the Hubble radius  in string frame $H_{s}^{-1}$ is
fundamentally different from the behavior of  the Hubble radius
in  the Einstein frame $H_{E}^{-1}$, as shown in Fig. 1, lines AB
versus DB. What if the dilaton was fixed due to some unspecified
nonperturbative processes at that time?  Then the universe
expanded in the same way in the Einstein frame and in the string
frame, and  $H_{E}^{-1}$ decreased in the same way as
$H_{s}^{-1}$ on line AB.

Of course, this would require stabilizing the dilaton not only at
the present stage of the evolution of the universe, which was not
achieved as yet in the framework of the model studied in
\cite{Nayeri:2005ck,Brandenberger:2006xi,Nayeri,Brandenberger:2006vv},
but also during the Hagedorn phase, and during the intermediate
phase between the Hagedorn regime and the radiation-dominated
epoch. We have no idea whether this could be done. As we have
already noted, the decrease of $H_{E}^{-1}$ would contradict the
null energy condition, which usually implies catastrophic vacuum
instability. With such dramatic nonperturbative changes to the
underlying physics, one can only guess what  happens to the
spectrum of density perturbations calculated in
\cite{Nayeri:2005ck,Brandenberger:2006xi,Nayeri,Brandenberger:2006vv}.
Moreover, if the dilaton is fixed at the time when the
cosmological perturbations are produced, then one can apply to
this situation the no-go theorem proven in \cite{mukhbook}, which
says that in this case the inflationary expansion is the only
mechanism which can generate cosmological perturbations with a
flat spectrum.

Interestingly, there is a direct relation between  the speculative
regime discussed above and inflation.  If one  manages to fix the
dilaton during the stage AB, when the cosmological metric
perturbations are supposed to be generated, then the Hubble
constants $H_{E}$ and $H_{s}$ are proportional to each other, and
both of them rapidly grow, so that ${dH_{E}\over d\tau}
> 0$. This means that
\be {d^{2}\alpha\over d\tau^{2}} = {dH_{E}\over d\tau} +
H_{E}^{2} > 0 \ , \ee
i.e. the universe in the Einstein frame is
accelerating.

The meaning of this result is very simple. If  $H_{E}$ is a
positive constant, or a slowly changing positive quantity, the
universe enters the regime of exponentially rapid accelerated
expansion, i.e. inflation. If $H_{E}$ is positive, and it grows in
time, then the rate of acceleration  also grows, so we  deal with
a super-exponential expansion, or super-inflation \cite{superexp}.
Note that we are not talking here about the contracting universe
which looks inflationary only in the string frame, which is the
case in the pre-big bang scenario. In this case the universe is
super-inflationary both in the string frame and in the Einstein
frame.

Thus, if we find some way to stabilize the dilaton in the Hagedorn
regime and remove the difference between the string frame and the
Einstein frame, which is one of the main stumbling blocks of the
scenario proposed in
\cite{Nayeri:2005ck,Brandenberger:2006xi,Nayeri,Brandenberger:2006vv},
then we end up in the inflationary universe anyway.

\section{Gravitational Waves}\label{grwaves}

Until now we have discussed scalar perturbations of the metric,
which play the key role in the cosmological structure formation. In this section
we turn to the generation of gravitational waves, which are tensor
perturbations.  We will critically review the derivation of
gravitational radiation in the Hagedorn thermal bath in the model
proposed in \cite{Brandenberger:2006xi}.  The mechanism of
gravitational waves radiation  from thermal bath is different from
that of gravitons production from vacuum in an expanding universe.

Gravitational waves perturbing the FRW background are the transverse-traceless (TT) metric fluctuations $h_{ij}$,
\begin{equation}
\label{TT} ds^2=-dt^2+a(t)^2 \, (\delta_{ij}+h_{ij}) \, dx^i dx^j
\, ,
\end{equation}
with $h^i{}_i=0$, $h^i{}_{j;i}=0$, $i,j=1,2,3$. The field equation
for gravitational waves is
\begin{equation}\label{motion}
\Box h_{ij}=\frac{16 \pi}{M_P^2} \, \pi_{ij} \ ,
\end{equation}
where $\pi_{ij}$ is the TT part of the energy-momentum tensor
$T_{\mu\nu}$. In the momentum space $\pi_{ij}$ is easily derivable
from $T_{\mu \nu}$ with the help of the projector operators
\begin{equation}
\label{anysotropy}
\pi_{ij}=\left(P_{il}P_{jm}-\frac{1}{2}P_{ij}P_{lm}\right) \,
T_{lm} \, ,
\end{equation}
where the projector operator is
\begin{equation}
\label{project}
P_{ij}(\hat k)=\delta_{ij}-\hat k_i \hat k_j \ , \,\,\, \hat k_i=\frac{ k_i}{|\vec k|} \ .
\end{equation}

If the
TT-part of energy-momentum tensor is absent, $\pi_{ij}=0$, then we
deal with the free gravity waves. This happens during inflation,
when the gravitational waves are generated from vacuum
fluctuations, even in the absence of any source $\pi_{ij}=0$.

Instead, let us consider now a simplified case of a static
universe filled with matter in thermal equilibrium. This was the regime
studied in \cite{Brandenberger:2006xi}. In this case gravitational
waves can be produced by the source term $\pi_{ij}$ even if the
universe is static. This mechanism, which is quite different from
the inflationary one, deserves a systematic study \cite{Ford,KU}.

First, let us
recall how this case was treated in \cite{Brandenberger:2006xi}.
It was suggested to use the Fourier-component version of the wave
equation (\ref{motion}),
\begin{equation}\label{mode}
k^2 h_{ij}=\frac{16 \pi}{M_P^2} \, \pi_{ij}(k) \, ,
\end{equation}
and construct the combination $k^3 |h_{ij}(k)|^2$ using this
equation. Here $k^2$ corresponds to the spatial three-momentum.
However, Eq. (\ref{mode}) is not a full Fourier-transform of
(\ref{motion}). An important term $\ddot h_{ij}(k)$ was dropped
from the equation. Thus Eq. (\ref{mode}) used in
\cite{Brandenberger:2006xi} is not a correct wave equation for the
radiation to be emitted by the sources.

Using the spatial Fourier-transform $h_{ij}(t, {\bf x}) \to
h_{ij}(t, {\bf k}) \, e^{i{\bf k x}}$, we can readily find the
correct version of Eq. (\ref{mode}). From (\ref{motion}) we have
\begin{equation}
\label{kmode} \ddot h_{ij}(k)+k^2 h_{ij}(k)=\frac{16 \pi}{M_P^2}
\, \pi_{ij}(t, k) \, .
\end{equation}
In a general  case of the FRW universe one finds a similar
equation, rescaling $h_{ij}\to \frac{1}{a} h_{ij}$ and switching
to the conformal time $\eta=\int dt/a(t)$:
\begin{equation}
\label{kmode1}
 h^{''}_{ij}(k)+(k^2 -a''/a) h_{ij}(k)=\frac{16 \pi}{M_P^2} \, a^3 \, \pi_{ij}(t, k) \, .
\end{equation}
For gravity waves generated inside the horizon, the term $ a''/a$
can be omitted.

Here we will only consider the limit of a static universe, in
order to compare our results with the conclusion of
\cite{Brandenberger:2006xi}. However, the  method of derivation of
gravitational radiation from the random media \cite{KU} which we
will use is quite general and can easily incorporate other cases,
of expanding/contracting universes, and of evolving dilaton. The
solution of equation (\ref{kmode1}) is
\begin{equation}
\label{osc} h_{ij}(k)=   \frac{16 \pi}{k\, M_P^2}   \int^t \,
dt'\, \sin k(t-t') \, \pi_{ij} (t', k) \, .
\end{equation}
The quantity $\pi_{ij}(t, k)$ along the characteristics of the
gravitational wave propagation is a random variable. We
can construct the variance of the gravity wave amplitude
\begin{eqnarray}
\label{amplitude} &&\langle |h_{ij}(t, k)|^2\rangle=\left(
\frac{16 \pi}{M_P^2} \right)^2 \, \frac{1}{k^2} \, \Bigl|\int^t \,
dt'\, \sin k(t-t') \nonumber\\ &\times&\int^t \, dt''\, \sin
k(t-t'') \langle \pi_{ij} (t', k)  \pi_{ij} (t'', k) \rangle\Bigr
| \, .
\end{eqnarray}
As we see, this expression is quite different from the naive
assumption of (\ref{mode}), and involves subtle information about
the time correlator $ \langle \pi_{ij} (t', k)  \pi_{ij} (t'', k)
\rangle$, which we can write  as
\begin{equation}
\label{corr} \langle \pi_{ij} (t', k)  \pi_{ij} (t'', k)\rangle
=f(t'-t'') \, \langle |\pi_{ij} (k)|^2\rangle \, .
\end{equation}
Let us take here the limit of large $t$. Replacing $\sin k(t-t')$
by exponents and dropping rapidly oscillating terms, one can
simplify Eq. (\ref{amplitude}). Formally, the result of the
calculations diverges with time, which is expected for infinitely
long stationary process of the gravitational waves production.
Indeed, nothing will interfere with the gravitational wave
production by colliding particles in the thermal gas, until the
gravitons come to a state of thermal equilibrium with matter. The
final result for  the gravitational wave production rate in a
thermal bath is
\begin{eqnarray}
\label{rate1} &&\frac{d}{dt} \langle |h_{ij}(t, k)|^2\rangle \\
&=&  \frac{(16 \pi)^{2}}{2k^2 M_P^4}    \, \langle |\pi_{ij}
(k)|^2\rangle \,\Bigl | e^{-ikt} \int^t \, dt'\, e^{+ikt'}
f(t'-t'') \Bigr | \ . \nonumber
\end{eqnarray}
To get an estimate of the spectrum of gravity waves generated in
string gas cosmology, we will take the estimate of
$\langle|\pi_{ij} (k)|^2\rangle$ given in
\cite{Brandenberger:2006xi}. They suggest that
\be \langle |\pi_{ij}|^2\rangle \simeq \frac{T(1-T/T_H)}{l_s^3
R^4} \ln^2\left[\frac{R^2}{l_s^2}(1-T/T_H)\right] \, , \ee
and so
\be \langle |\pi_{ij} (k)|^2\rangle \simeq  k \,
\frac{T(1-T/T_H)}{l_s^3 }
\ln^2\left[\frac{1}{k^2l_s^2}(1-T/T_H)\right] \, . \ee

The time correlation function of thermal fluctuations  $f(t'-t'')$
should be a decreasing function of time separation. Let us suppose
that $f(t'-t'') \sim e^{-T (t'-t'')}$. Then from (\ref{rate1}) we
obtain the final estimate
\begin{eqnarray}
\label{final} &&k^3 \langle |h_{ij}(t, k)|^2\rangle \simeq
\frac{(16 \pi)^{2}k^2}{2 \sqrt{k^2+T^2} M_P^4} \nonumber\\
&\times& \frac{T(1-T/T_H)}{l_s^3 }
\ln^2\left[\frac{1}{k^2l_s^2}(1-T/T_H)\right] \, \Delta t
\, ,
\end{eqnarray}
where $\Delta t$ is the duration of the process. The meaning of
this result is that the gravitational waves are constantly
generated as long as we have a thermal bath containing many
moving and colliding particles. In an expanding/contracting
universe $\Delta t$ should be replaced by the age of the universe
$\sim 1/H$. The spectrum is not scale-free but peaked at the scale
of temperature $k \sim T$, as expected for the radiation produced
by matter in thermal equilibrium.

This shows how to consistently improve the results of Ref.
\cite{Brandenberger:2006xi}.
As we have seen, these calculations describe production of
gravitational waves by
 thermal fluctuations
 assuming that the
gravitons are out of thermal equilibrium, and yield a spectrum
which is not flat.
The state of the universe in which our calculation applies simply
doesn't contain many gravitons. This is certainly the case at the
late stages of the evolution of the universe, but this
should not be the case if the initial state was a Hagedorn phase
in thermal equilibrium. If this equilibrium
initial state is possible, then the gravitons are constantly
emitted and absorbed, and their spectrum cannot be flat.

Note that the gravitons correspond to the lowest energy
excitations of the closed string loops. The existence of the
Hagedorn temperature is a consequence of the existence of a tower
of massive states giving large contribution to the total energy
density. When the temperature drops below the Hagedorn
temperature, the gravitons are produced not only by the thermal
radiation studied above, but also due to the decay of the massive
string excitations, i.e. due to the decay of large strings.  On
top of that, one should take into account the dilaton dynamics and
the difference between the string frame and the Einstein frame,
which leads to many important consequences discussed in the
previous section.
These questions have not been analyzed in
\cite{Nayeri:2005ck,Brandenberger:2006xi,Nayeri,Brandenberger:2006vv},
and we won't pursue this issue here any further either.

\section{Discussion}

Our investigation shows once again that it is very difficult to
offer a non-inflationary solutions of the major cosmological
problems.
It remains quite instructive, however, to learn why is it so
difficult to find an alternative to inflation. In this particular
case, the difficulties start from the very beginning, since even
the existence of the basic cosmological solution is rather
problematic, see Section \ref{basic}.
Nevertheless, one could still
nurse a hope that this situation could somehow be improved in
future development of the model, by adding new ingredients,
finding a better effective action, etc.

However, we
do not think that this could help.
Our main concern stems from the fact that one of the ingredients
of production of the cosmological perturbations in the scenario
proposed in \cite{Nayeri:2005ck,Nayeri,Brandenberger:2006vv} is
purely kinematic: Perturbations should freeze when they cross the
horizon. This,
of course, is the main idea of the inflationary mechanism of
generation of perturbations of metric: during inflation the
wavelength of the inflaton perturbations grow, and when the
wavelength becomes greater than the size of the horizon, the
perturbations freeze. The new idea of Ref.
\cite{Nayeri:2005ck,Nayeri,Brandenberger:2006vv} is that in their
model the wavelengths of the perturbations grow very slowly, but
the quantity $H_{s}^{{-1}}$
should rapidly shrink. Then,
they posit, perturbations freeze when
$H_{s}^{{-1}}$ becomes smaller than the wavelengths of the
perturbations. However, this was just a conjecture, based on an
analogy with the results obtained in inflationary cosmology.

The easiest way to
test this conjecture in the model proposed in
\cite{Nayeri:2005ck,Brandenberger:2006xi,Nayeri,Brandenberger:2006vv}
is to study the evolution of the perturbations in the Einstein
frame. In this frame, the perturbations shown in Fig. 1 from the
very beginning have the wavelength
much greater than $H_{E}^{{-1}}$, so the analogy with the
inflationary mechanism breaks down. This is one of the reasons why
the spectral index of the cosmological perturbations generated in this scenario is unacceptably large, $n = 5$.

Of course, it might happen that eventually one will find a version
of this model where the size of the horizon $H_{E}^{{-1}}$ evolves
extremely slowly, so that for a long time it remains nearly
constant. Then the cosmological evolution will cause the
wavelengths of perturbations to cross the horizon, which, under
certain conditions, may lead to their freezing. But the slow
evolution of the Hubble constant $H_{E}$ is a definitive feature
of inflationary cosmology. Thus, at the moment we do not see any
reason to expect that this model or its possible generalizations
can provide a viable alternative to inflation.

{\bf Note Added:} The main results of this paper were reported and
debated with the authors of
\cite{Nayeri:2005ck,Brandenberger:2006xi,Nayeri,Brandenberger:2006vv}
during the conference ``Inflation+25'', Paris, June 2006. Recently
Brandenberger, Nayeri, Patil and collaborators issued a paper
\cite{Brandenberger:2006pr} where they confirmed our statement
that the perturbations produced in the original scenario of
\cite{Nayeri:2005ck,Brandenberger:2006xi,Nayeri,Brandenberger:2006vv}
have an unacceptably large spectral index $n = 5$. They argued
that this scenario can be salvaged if the dilaton in the early
universe were fixed  by some unspecified strong coupling effects.
We discuss this speculative possibility in Sect.  \ref{para}, and
conclude that it violates the null energy condition. Even if such
regime were possible, we show that the universe in this regime
would accelerate both in the string frame and in the Einstein
frame. In other words, the universe in this speculative regime is
inflationary, so it does not provide a true alternative to
inflation.

\acknowledgements

The authors are grateful to N.~Arkani-Hamed, R.~Brandenberger,
D.~A.~Easson, R.~Kallosh, A.~Mazumdar, A.~Nayeri, M.~Sasaki, A.~Starobinsky,
A.~Tseytlin, C.~Vafa and S.~Watson for many useful discussions. NK
would also like to thank Galileo Galilei Institute, Florence,
Italy, and to LPT, Universite de Paris-Sud, Orsay, France, for
kind hospitality during the course of this work. The work of NK
was supported in part by the DOE Grant DE-FG03-91ER40674, by the
NSF Grant PHY-0332258 and by a Research Innovation Award from the
Research Corporation. The work of L.K. was supported by NSERC and
CIAR. The work of A.L. was supported by NSF grant PHY-0244728 and
by the Humboldt Foundation.

\end{document}